\documentclass{ledger}

\newcommand{\thefirstpagenum}[0]{X}
\newcommand{\thelastpagenum}[0]{X}
\newcommand{\theyear}[0]{20XX}
\newcommand{\thevol}[0]{X}
\newcommand{\thedoi}[0]{DOI 10.5195/LEDGER.\theyear.XXX}
\newcommand{\ledgerpages}[0]{\thefirstpagenum-\thelastpagenum}

\hypersetup{pdfauthor={Amherd, Lucas; Li, Sheng-Nan; Tessone, Claudio J}, pdftitle={Centralised or Decentralised? Data Analysis of Transaction Network of Hedera Hashgraph}}

\title{Centralised or Decentralised? Data Analysis of Transaction Network of Hedera Hashgraph}

\author{Lucas Amherd, \thanks{L. Amherd (lucas.amherd@uzh.ch) is a member of the Faculty of Business, Economics and Informatics at the University of Zurich, Switzerland} Sheng-Nan Li, \thanks{S.N. Li (shengnan.li@uzh.ch) is a researcher at Blockchain and Distributed Ledger Technologies group at the University of Zurich and the UZH Blockchain Center} 
Claudio J. Tessone \thanks{C. J. Tessone (claudio.tessone@uzh.ch) heads the Blockchain and Distributed Ledger Technologies group at the University of Zurich. He is also co-founder and Chairman of the UZH Blockchain Center. }}

\pretitle{
  \centering \selectfont RESEARCH ARTICLE \par 
  \fontsize{24pt}{28pt}\selectfont} 
  
\begin{document}

\maketitle

\thispagestyle{pagefirst}

\begin{abstract}
An important virtue of distributed ledger technologies is their acclaimed higher level of decentralisation compared to traditional financial systems. Empirical literature, however, suggests that many systems tend towards centralisation as well. This study expands the current literature by offering a first-time, data-driven analysis of the degree of decentralisation of the platform Hedera Hashgraph, a public permissioned distributed ledger technology, employing data directly fetched from a network node. The results show a considerably higher amount of released supply compared to the release schedule and a growing number of daily active accounts. Also, Hedera Hashgraph exhibits a high centralisation of wealth and a shrinking core that acts as an intermediary in transactions for the rest of the network. However, the Nakamoto index and Theil index point to recent progress towards a more decentralised network.

\begin{keywords}
\item Distributed Ledger Technology.
\item Cryptocurrency.
\item Hedera Hashgraph.
\item Transaction Network.
\item Decentralisation.
\end{keywords}
\end{abstract}

\section{Introduction}\label{chapter:introduction}
The beginning of a new era unfolded in 2008 when Satoshi Nakamoto, a group or individual operating under a pseudonym, published the Bitcoin whitepaper, which proposed a novel solution to the traditional centralised payment system by resolving for the first time the double spending problem \cite{Nakamoto2008}. To prevent users of digital currency from spending the same currency or token twice, Nakamoto introduced a technology that had been in existence for some time: the blockchain. It is a digital, append-only, and distributed ledger secured through cryptography and modified using a consensus protocol. A distributed ledger is a store of data spread across many geographical places and is updated via a consensus algorithm. Numerous distributed ledger technology (DLT) systems aim to achieve decentralised secured by cryptography, commonly referred to as "cryptocurrencies", following Bitcoin's example as a peer-to-peer system. \par

Decentralisation is a crucial component of all distributed ledger technologies, including blockchain and hashgraph technology \cite{puthal2018blockchain,weyl2022decentralized}. The exact reason for their emergence is the increasingly centralised state of the more and more digitalised economy which we trust, leading to fewer institutions possessing more power. An example is Bitcoin, where the first block in its history, the Genesis Block, contained the headline of a British newspaper: \emph{"The Times 03/Jan/2009 Chancellor on brink of second bailout for banks"} \cite{TheTimes}. This is often interpreted as a call for people to stop trusting centralized banks and instead begin to rely on a more decentralized system. Consequentially, decentralisation is seen as an integral virtue of distributed ledger technology (DLT) systems. However, some literature has shown that these systems tend towards centralisation as well \cite{beikverdi2015trend, gencer2018decentralization, campajola2022evolution, budish2022economic, sultanik2022blockchains, gupta2018gini, gochhayat2020measuring, cong2023inclusion}. Therefore, it is crucial to differentiate between the common narrative and the empirical evidence regarding the level of centralization in DLT systems. \par

Currently, there is a dearth of quality literature on Hedera Hashgraph (subsequently also called "the network"), particularly when it comes to its decentralisation, which is not surprising given that the network launched only in 2018 and  has not yet achieved the size of the market capitalization of major cryptocurrency platforms like Bitcoin and Ethereum \cite{Buterin2014}. In one paper, Krasnoselskii et al.\cite{krasnoselskii2020distributed} proposed a random number generator solution implemented on Hedera Hashgraph, while Jam et al.\cite{naser10hungerhash} described a charity network partly implemented on Hedera Hashgraph. Previous works focus more on the implementation, while this work seeks to contribute to the scientific literature on the Hedera network.\par

Conceptually, decentralisation does not always have a standard definition; Vitalik Buterin discusses architectural, political and logical decentralisation \cite{buterin2017meaning}. Halaburda and Mueller-Bloch\cite{halaburda2020toward} and Karakostas, Kiayias, and Ovezik\cite{karakostas2022sok} propose a systematic, multiple-layer approach to evaluating decentralisation in DLTs. A different work by Howell, Saber, and Bendechache\cite{howell2023measuring} quantifies node decentralisation not only geographically but also the number of validator nodes and their running software versions for which they propose the NodeMaps framework. In another paper, while highlighting the challenges of eliciting the level of decentralisation of a DLT system, Lee et al.\cite{lee2021dq} calculate a decentralisation quotient of geographical diversity and censorship resistance of the data. Complementary work also comes from the social and economic perspective \cite{bodo2019logics,bodo2021decentralisation}. Sai et al. \cite{sai2021taxonomy} based on an analysis of 89 research papers from which they propose a centralisation taxonomy, while a different taxonomy offered by Zhang et al. \cite{zhang2022sok} is based on consensus, governance, network, transactions and wealth. These few examples show that the concept of decentralisation in the context of DLT systems can be defined very differently. Since the structure of a DLT system, which consists of entities transacting among each other, can be easily represented in the form of a network, this paper methodologically follows chiefly Campajola et al.\cite{campajola2022evolution} employing various concepts from complex systems literature such as degree distributions and the core-periphery structure to determine the degree of decentralisation. \par

The purpose of this work is to analyse Hedera Hashgraph on its level of decentralisation and to contribute first-time results from a proof-of-stake (PoS) based distributed ledger technology to the literature by generalising the results in existing literature\cite{bovet2019evolving, NBERw29396, campajola2022evolution, kusmierz2022centralized, lin2021measuring, gochhayat2020measuring}. Another distinctive feature this network exhibits consists of the hashgraph structure that compares to the much more common blockchain technology. This paper also provides new results in this subdomain of DLT systems. 
The research question of this study is formulated as follows: \emph{"What is the level of centralisation exhibited by the Hedera Hashgraph token transaction network according to different centralisation measures?"} 
\par

As the following structure of this paper: the protocol of Hedera Hashgraph will be introduced briefly in \cref{chapter:Hedera Hashgraph}. In \cref{chapter:data and methodology} the data structure and methods used to analyse Hedera Hashgraph are presented. The chapter \cref{chapter:results} shows a high-level analysis of the activity on Hedera Hashgraph including the number of transactions over time or the rate of active accounts over time. Moreover, results on various decentralisation measures such as the Gini coefficient on wealth distribution or the core-periphery structure of the network are presented in \cref{chapter:results}. The discussion in \cref{chapter:discussion} concludes the analysis.

\section{Hedera Hashgraph}\label{chapter:Hedera Hashgraph}
\subsection{The Hashgraph Protocol}\label{section:Hashgraph protocol}
This paper focuses on the token transaction network of Hedera Hashgraph, a distributed ledger technology controlled by a network of computers, not a single entity. The network runs on the Hashgraph protocol invented by Leemon Baird \cite{baird2020hashgraph}, which records transactions in a directed acyclic graph (DAG) structure, rather than using blockchain, the more common distributed ledger technology. The Hashgraph protocol is an asynchronous Byzantine Fault Tolerant protocol (aBFT) that ensures the network's functioning in the presence of malicious nodes, including time-based attacks such as distributed denial of service (DDoS) attacks. 
The entire transaction and communication history between nodes - called the hashgraph - is stored by all the Hedera Council Member nodes, currently a select set of diverse institutions, and broadcasted to the network. Transactions in the network are spread by gossip, where each node that received a transaction selects another node at random and synchronizes its hashgraph with that node. An in-depth description of the Hashgraph protocol is out of scope for this work and provided by Sridhar, Blum, and Katz\cite{sridhar2022musings}. \par

\subsection{Permissioned and Permissionless}\label{section:approach to decentralisation}
Hedera Hashgraph was launched in August 2018 as a public permissioned DLT platform, whereas public refers to the fact that anyone can make transactions and permissioned to the fact that nodes operating the network are selected upon specific criteria. Hedera Hashgraph LLC. is currently governed by the Hedera Governance Council (HGC) members \cite{hedera_decentralisation_new}. HGC members are selected to be as diverse as possible regarding the type of institution, the geographical location and the industry in which they operate. Furthermore, they should be large institutions with an international reputation to make the governance more credible \cite{hederacouncil}. The HGC takes the role of several committees (Membership committee, technical steering and product committee, corporate utilization committee, coin economics committee and legal and regulatory committee) and manages the treasury of the network. \par

When it comes to decentralisation, Hedera Hashgraph will take a methodical approach consisting of three phases: In phase 1, the network is run and transactions are verified only by consensus nodes operated by the members of the HGC, a diverse set of institutions each having an equal influence on the network. Then, phase 2 will allow more permissioned consensus nodes to secure the network by verifying transactions without being part of the HGC. By the time hundreds of permissioned consensus nodes run on the network and the HGC has reached the maximum number of 39 members, Hedera Hashgraph will transition to a \emph{permissionless} system, enabling anyone to run anonymously a consensus node and earning rewards in the form of the native currency HBAR for securing the network (for more detail see \ref{section:network supply}). A native currency is the base currency of a DLT system, which is for instance used to pay transaction fees. This approach will be further elaborated in \cref{chapter:results} and \cref{chapter:discussion}. \par

\section{Data and Methodology}\label{chapter:data and methodology} 
This section provides a brief overview of the employed data and an introduction to the metrics used in assessing the level of decentralisation in the subsequent analyses in chapter \ref{chapter:results}. 
\subsection{Data Set}\label{section:data base}
The analysis is based on data collected directly from the Hedera Hashgraph network via its mirror node using the REST Application Programming Interface (API). 
The REST API provides access to information on various ledger variables, including accounts, transactions, balances, tokens or smart contracts. Starting from 13 September 2019, we collect weekly balance data and all transaction history until and including 28 February 2023, a total of 3 years, 5 months and 16 days. Despite the network having already launched on 24 August 2018, the mirror nodes only started storing a continuous hashgraph history as of 13 September 2019. The tables and figures presented in chapter \ref{chapter:results} depict the results of this time period. \par 

\begin{figure}[h]
    \centering
    \captionsetup{justification=centering}
    \includegraphics[scale=0.4]{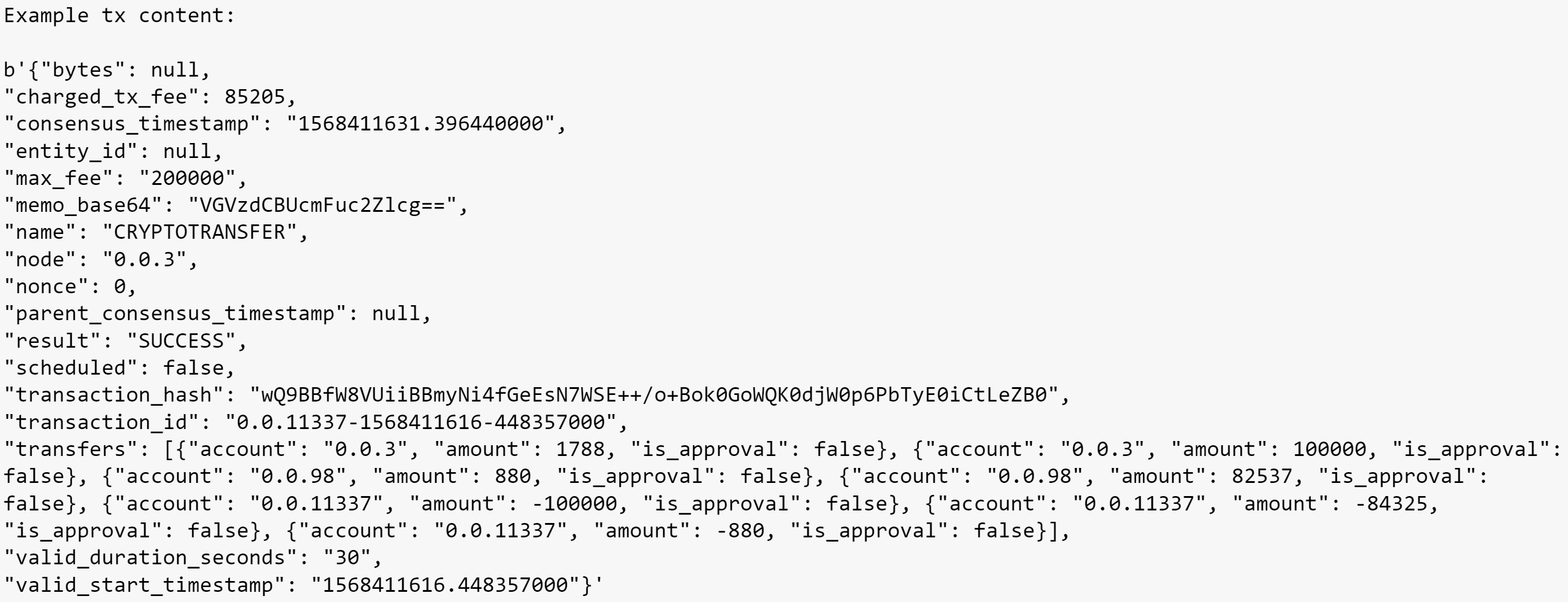}
    \caption{Example of one transaction in the raw data format.}
    \label{fig:example_tx}
\end{figure}
An exemplary structure of the transaction as it is fetched from the mirror node is presented in figure \ref{fig:example_tx}. The variable "consensus timestamp" marks the time when consensus was reached on the transaction in Unix epoch format. The variable used to identify the transaction type, such as HBAR transfer, token transfer or account creation, is "name". Most importantly, the variable "transfers" lists all accounts involved in the transaction, including the respective change in their balance ("amount"). \par 
The account-based model used in Hedera Hashgraph is a data accounting standard where private-public key cryptography manages the access rights to the assets in an account. 
To note, the wealth of accounts does not necessarily equal the wealth of a user and vice versa since one user may have multiple accounts, and one account may include the wealth of multiple users, it is challenging to accurately identify the number of users. Nonetheless, in this paper, specific accounts are labelled in the network explorers, and Hashscan, developed by Swirlds Labs, is used to facilitate this identification process.

\subsection{Gini Coefficient}\label{section:desc Gini coefficient}
The Gini coefficient is a simple way to gauge inequality of different kinds. It is derived from the Lorenz curve which states what share of, for instance, wealth is possessed by what share of the population. 
If there are $N$ entities and $x_i$ is the wealth of entity $i$, then the Gini coefficient defined as:

\begin{equation}
    G=\frac{\sum_{i=1}^{N}{\sum_{j=1}^{N}{|x_i-x_j|}}}{2N\sum_{i=1}^{N}{x_i}}
\label{eqG}    
\end{equation}

It is therefore in the interval of [0,1], one corresponding to complete \emph{inequality} where one individual has all the wealth and all the others possess nothing.

\subsection{Degree Assortativity}\label{section: desc degree assortativity}
In the section degree distribution \ref{section:degree distribution}, the node degree is defined as the number of other nodes a focal node transacts with, and in the case of an account-based model, a node is represented by an account. A high degree could be translated to a high importance of a node in the network. Furthermore, the degree assortativity is a measure to quantify the extent to which high-degree nodes connect with other high-degree nodes (assortative) or low-degree nodes (disassortative). The assortativity coefficient is normalised to [-1,1] where a negative coefficient means that the network is disassortative. 

When $r$ is the assortativity coefficient, it is given by,

\begin{equation}
    r = \frac{M^{-1} \sum_{i}^{M} j_{i} k_{i} - [M^{-1} \sum_{i}^{M} \dfrac{1}{2}(j_{i}+ k_{i})]^2} {M^{-1} \sum_{i}^{M} \dfrac{1}{2}(j_{i}^2+ k_{i}^2) - [M^{-1} \sum_{i}^{M} \dfrac{1}{2}(j_{i}+ k_{i})]^2},
 \label{eq:introduction:r}
\end{equation}
where $j_i$ and $k_i$ are the degrees of the nodes at the ends of the edge $i$, and M is the total number of edges \cite{liang2018evolutionary} .

\subsection{Core-Periphery Structure}\label{section:desc core-periphery structure}
When it comes to economic networks or social network analysis in general, they often exhibit a "core-periphery structure" \cite{barucca2016disentangling, lip2011fast, Smith1992}. The idea is that the actors can be partitioned into two distinct groups, so-called subgraphs. One is the "core", in which all the actors are densely connected within each other, and the "periphery", in which the actors are chiefly connected to the core and only sparsely among each other. For the calculation this paper follows the algorithm presented by Lip \cite{lip2011fast}.

\subsection{Nakamoto Index}\label{section:desc nakamoto index}
The Nakamoto index version employed here, similar to Lin et al.\cite{lin2022weighted}, quantifies the minimum number of entities that need to collude in order to successfully conduct a Sybil attack, which allows them to control the network \cite{srinivasan2017quantifying}. The higher the Nakamoto index, the more resilient and secure a network is. On Hedera Hashgraph, to run a successful Sybil attack, one-third of the voting power is required, which is achieved by holding one-third of the released HBAR supply, the network's native coin, under the assumption that the consensus is formed on a one-HBAR, one-vote basis \cite{hbareconomics}. It is expressed as:

\begin{equation}
    N_c=  \frac{1}{n} min\{k \in \left[1,2,...,n\right] : W^{-1} \sum_{i=1}^{k}{w_i} > s\}
\label{eqNc}    
\end{equation}
where $w_i$ is the wealth of entity $i$ and there are again $n$ entities. The entities are sorted by descending wealth, i.e.~$w_i \leq w_j$ if $i \geq j$. Two values are used in the results figure \ref{fig:nakamoto index} for $s$, the supply share required: $1/3$ and $1/2$.

\subsection{Theil-T Index}\label{section:desc theil index}
Another statistic for decentralisation, the Theil-T index, can be borrowed from the field of information theory. It is an entropy-based gauge chiefly used for economic inequality and can be interpreted as the degree of non-randomness or redundancy \cite{theil1972statistical}. Mathematically, if $x_i$ is the wealth of entity $i$ with a total of $N$ entities, then it is given by,
\begin{equation}
    T_{T}=\frac{1}{N}\sum_{i=1}^{N}{\frac{x_i}{\mu}\ln{\left(\frac{x_i}{\mu}\right)}}
\label{eqTt}    
\end{equation}

Thus, the Theil-T index measures in our case how far the network is from equal distribution of wealth. A value of 0 corresponds to a completely egalitarian state of the distribution, whereas the maximum entropy means one account has all the wealth. 

\section{Results}\label{chapter:results}
This section starts with a general overview of the token transaction network in the form of various measures, such as the number of transactions and the daily active accounts over time, as well as the evolution of network supply, which are derived from the analysis of transaction data. It continues by providing gauges for decentralisation, namely the Gini coefficient of wealth, the degree distribution, the core-periphery structure, the Nakamoto index and the entropy. \par

\subsection{Number of Transactions}\label{section:tx}

\begin{figure}[h]
    \centering
    \captionsetup{justification=centering}
    \includegraphics[scale=0.45]{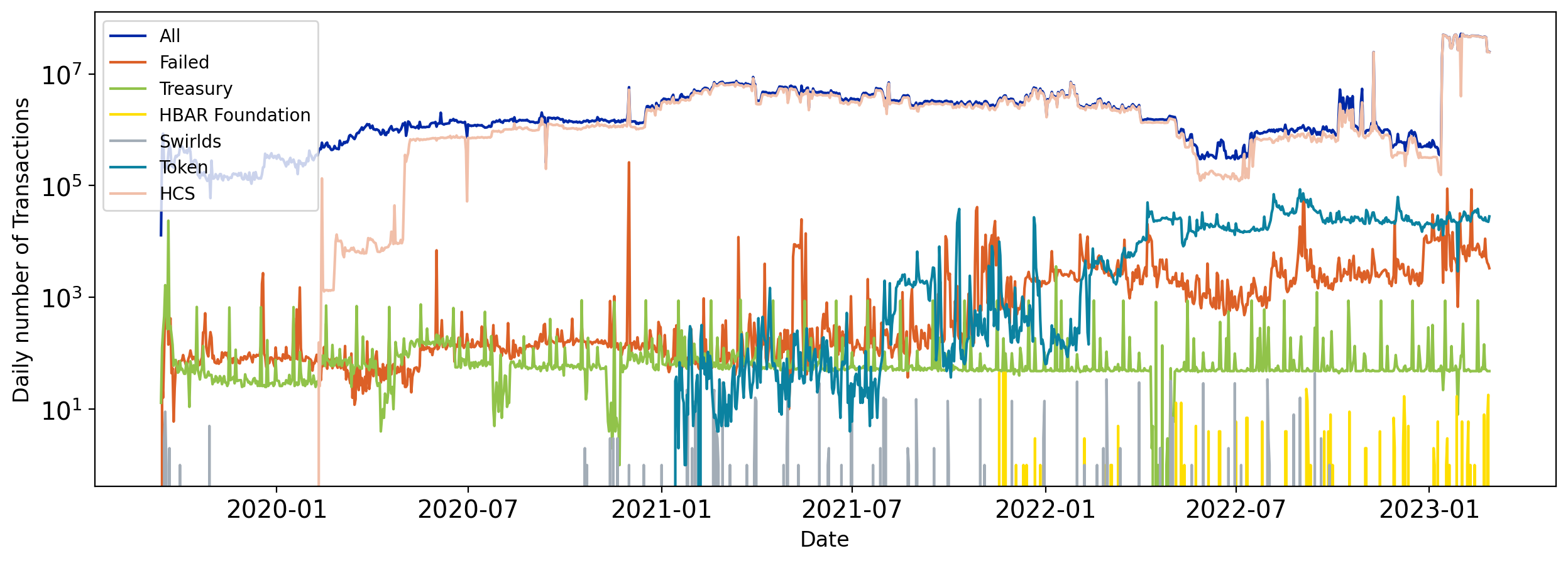}
    \caption{Number of transactions over time in absolute numbers, categorized by type.}
    \label{fig:transaction_num}
\end{figure}

The most basic way to measure network activity is the daily number of transactions. In Figure \ref{fig:transaction_num}, all transactions are shown along with subcategories. Unsuccessful transactions only constitute a small and stable fraction of all transactions, whereas the categories "Treasury", "Hbar Foundation" and "Swirlds" (Swirlds Labs) are the three identifiable user groups in this network. \newline

\begin{figure}[h]
    \centering
    \captionsetup{justification=centering}
    \includegraphics[scale=0.6]{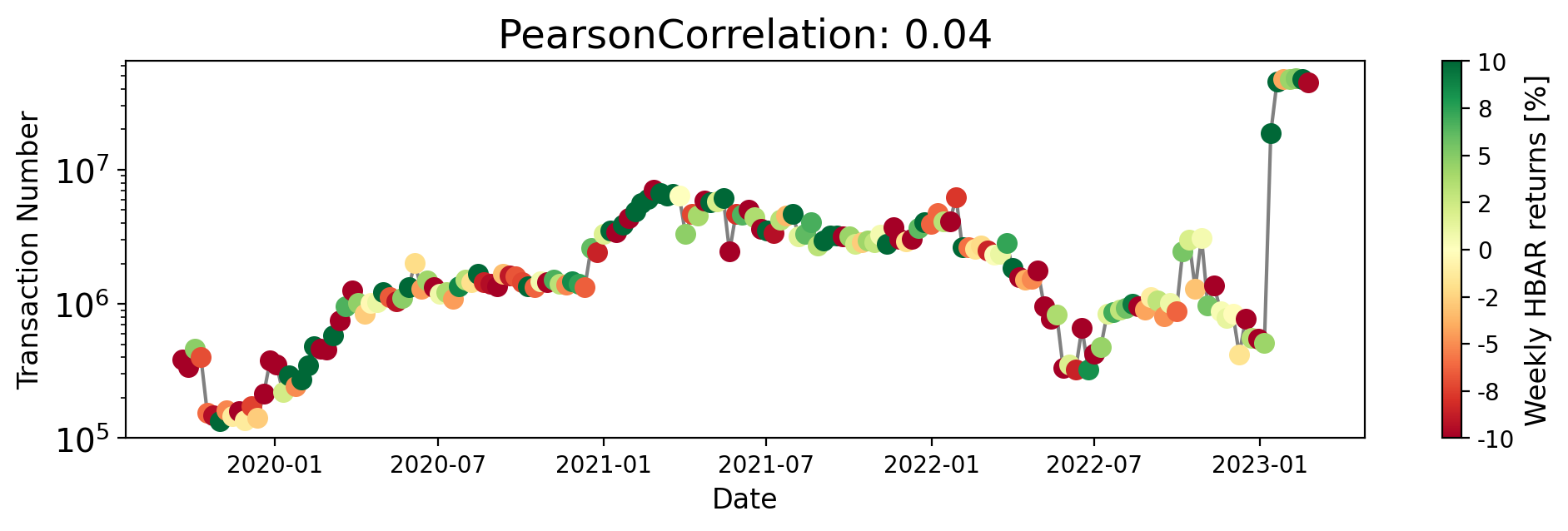}
    \caption{Evolution of daily number of transactions colour-coded with the weekly HBAR returns.}
    \label{fig:hbar-txnum_corr}
\end{figure}

As we can see in figure \ref{fig:transaction_num}, the vast majority of transactions are  from the Hedera Consensus Service (HCS). Moreover, the jump in total transactions in January 2023 by 140x within 3 days is worth to be mentioned. The spike occurred from increased demand in HCS, introduced on the 10th of February 2020, and hence led to a decrease in the percentage of token and treasury transactions. This figure also shows that the Treasury is involved in around 100 daily transactions, as opposed to the Hbar Foundation and Swirlds which merely make occasional transactions \ref{fig:transaction_num}. However, noticeable is the trend of the increasing number of token transactions, i.e. non-HBAR transactions, starting from the 14th of January 2021 when the Hedera Token Service was introduced on the Hedera Hashgraph mainnet and reaching around 5\% in mid-2022 where also the number of token transactions plateaus (see Appendix A: figure \ref{fig:transaction_num_pct}). \newline

Another interesting finding from figure \ref{fig:hbar-txnum_corr} is that the value of the network's native token, HBAR, has a very weak correlation with network activity and is highly correlated with the value of the dominant cryptocurrency Bitcoin (Pearson correlation of 0.917 over the analysed time period, see Appendix A: figure \ref{fig:hbar-btc_corr}) which suggests that the price of HBAR does not reflect the efficient price of the network at all points in time, i.e. is not evaluated according to the network-specific information, but is rather traded along with the market dominant asset Bitcoin. \newline

\subsection{Active Accounts}\label{section:active accounts}
\begin{figure}[h]
    \centering
    \includegraphics[scale=0.53]{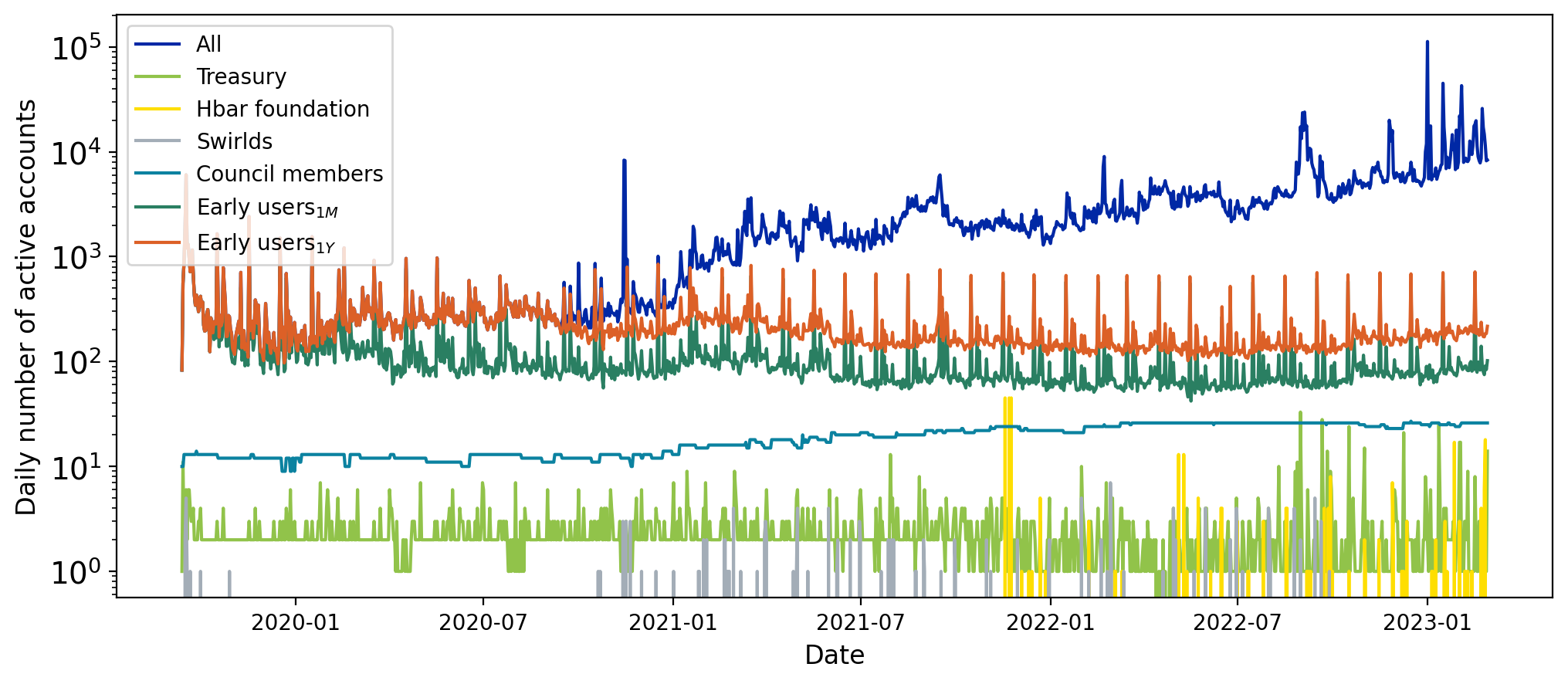}
    \includegraphics[scale=0.3]{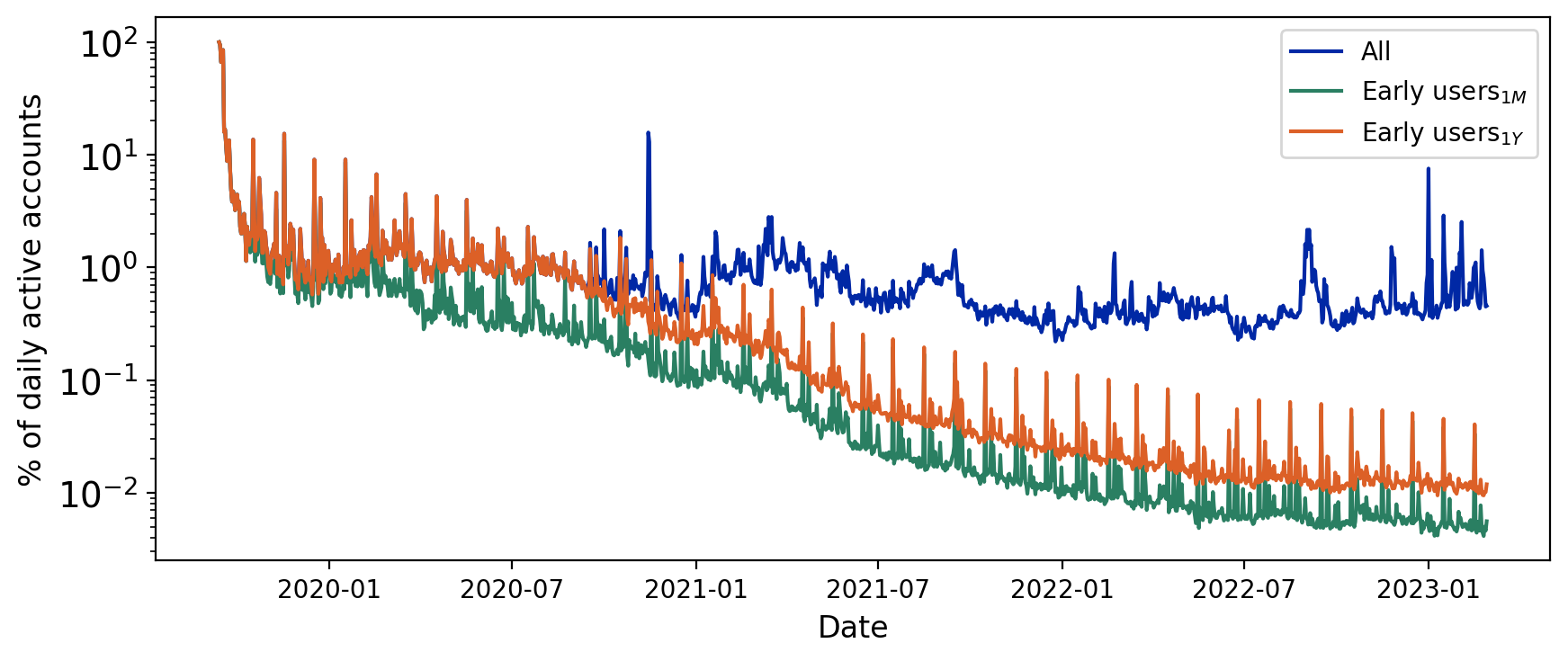}
    \includegraphics[scale=0.3]{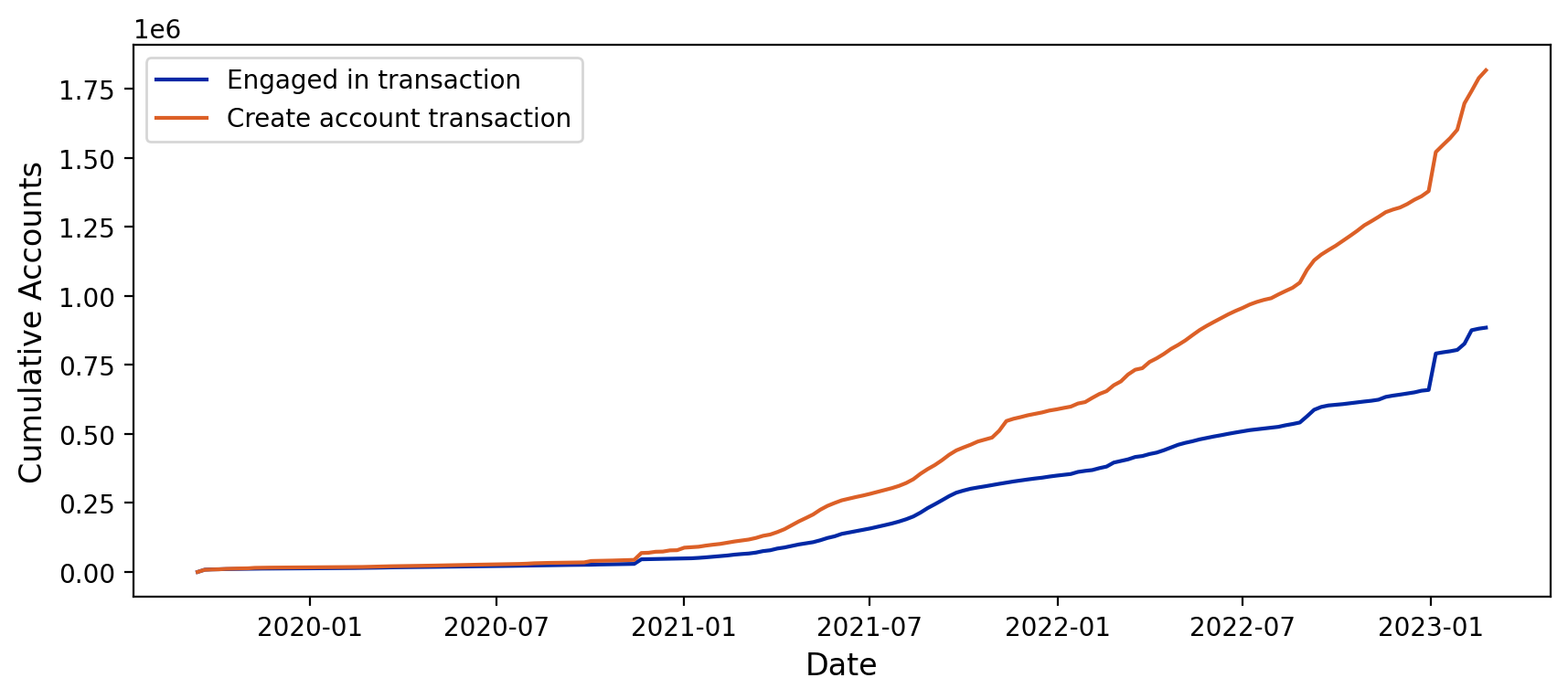}
    \caption{Daily Active Accounts in absolute numbers (top), as a percentage of the cumulative accounts (bottom left). Cumulative accounts over time (bottom right).} 
    \label{fig:active accounts}
\end{figure}

In figure \ref{fig:active accounts}, the subcategories "Early users$_{1M}$" and "Early users$_{1Y}$" refer to accounts that were created (also called born) during the first month (1M) or first year (1Y) of the analysed period, i.e. engaged in a transaction for the first time in this time period. For instance, the subcategory "Early users$_{1M}$" consists of all accounts that engaged for the first time in a transaction between 13.09.2019 and 13.10.2019, and it is a subset of "Early users$_{1Y}$". The main insight from figure \ref{fig:active accounts} is that the total number of daily active accounts in the Hedera transaction network is steadily growing from a few hundred to a few ten thousand; i.e. since its launch, it has increased by two order of magnitude. 
Focusing on the early users, a periodical pattern can be observed on a monthly basis around the 16th and 17th day of the month. In addition, the bottom left figure in Figure \ref{fig:active accounts} shows that the percentage of daily active accounts among the total accounts has a decreasing trend and lies mostly below 1\% in 2022, which is in stark contrast to the growing number of active accounts, pointing at a relatively higher growth rate of total accounts in bottom right figure of Figure \ref{fig:active accounts}. \par

\subsection{Released Supply}\label{section:network supply}
\begin{figure}[h]
    \centering
    \captionsetup{justification=centering}
    \includegraphics[scale=0.7]{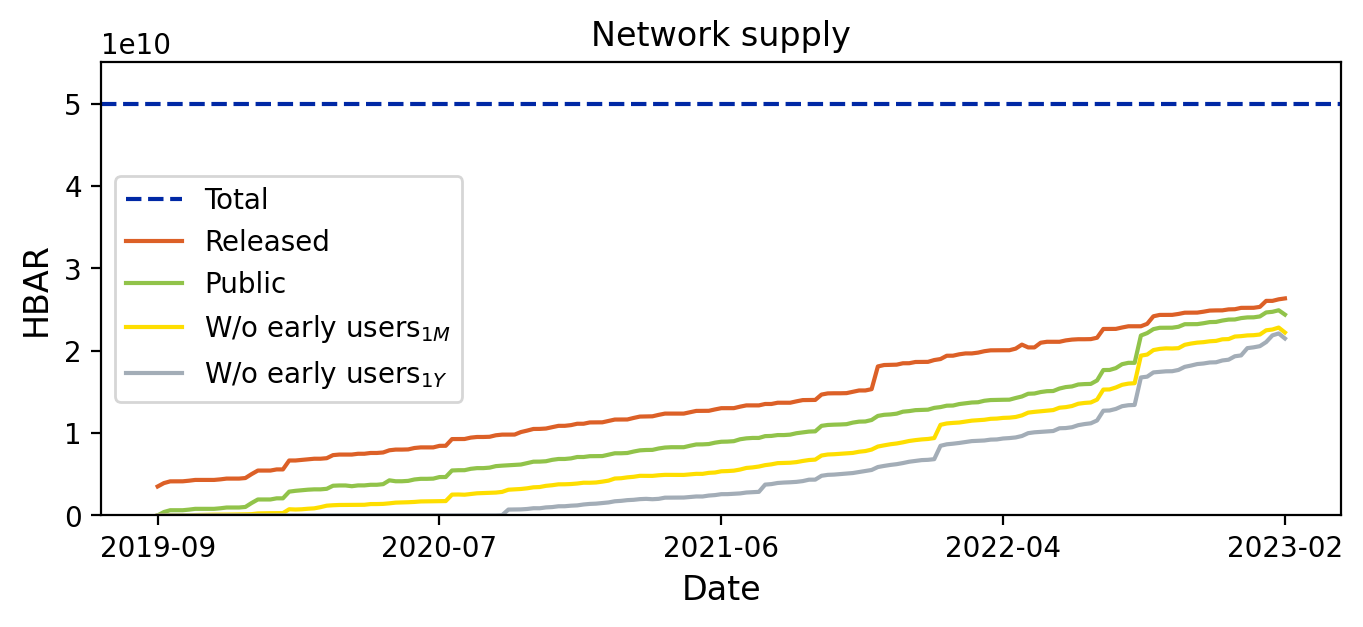}
    \caption{Released supply on Hedera Hashgraph with maximum supply at 50 billion HBAR.}
    \label{fig:supply}
\end{figure}

The past distributed or released Hbar supply to the public 
is displayed in figure \ref{fig:supply} and exhibits a nearly linear increase over time. The goal of the treasury is to achieve a wide distribution of the coins in order to ensure network safety once it has transitioned to a permissionless model (as mentioned in \ref{chapter:introduction}). Network safety is determined by the coin distribution, as the amount of HBAR represents the proportional amount of voting power in its PoS consensus mechanism.

A contradiction occurs from their official release schedule where they claim the released coins would not surpass 34\% of the total supply until 2025, however, this threshold was already passed in late November 2021 \cite{hbareconomics}. This might be due to the fact that the coin distribution was more even than expected so that further coins could be released without relinquishing much of the wealth distribution or that, in a more pessimistic interpretation, the coins needed to be sold to finance the continuing development of the network. 
The converging trend of the released supply and the balance among all accounts born after the first year shows the relatively strong flow of tokens to newly created accounts. 

\subsection{Degree Distribution}\label{section:degree distribution}

As Campajola et al. \cite{campajola2022evolution} point out, many economic phenomena and  financial systems, e.g. the Italian interbank market \cite{bargigli2015multiplex}, exhibit a strongly skewed distribution, similar to a power-law Pareto type distribution ($\alpha$ around 2.3). Their statistics on the $\alpha$ exponent of a Pareto fit of the degree distribution over several proof-of-work based (PoW) cryptocurrencies, such as Bitcoin, Ethereum (pre-PoS-merge), Monacoin and Litecoin, are highly identical to the results on Hedera Hashgraph (HBAR) and are reported in table \ref{t:power-lawFit}. For example, while the Hedera Hashgraph network has a median $\alpha$ of 2.11, Campajola et al.\cite{campajola2022evolution} find values of 2.69 for Bitcoin (UTXO-based) and 1.76 for Ethereum (Account-based). Their remark that the resemblance observed across various cryptocurrency systems suggests that these transaction networks possess a scale-free structure (many nodes with few connections and few highly-connected nodes (hubs)) is supported by the findings on Hedera Hashgraph. The findings on Hedera Hashgraph degree distributions also give a new insight into degree distributions of PoS DLT systems, which yield no huge difference to PoW systems. This finding has, of course, to be confirmed by more evidence to give it more external validity. \newline
\begin{table}[h]
    \centering
    \captionsetup{justification=centering}
        \begin{tabular}{ p{1cm}|p{2cm}|p{1cm}|p{1cm}|p{1cm}|p{1cm}}
        & \multicolumn{1}{c|}{} & \multicolumn{1}{c|}{HBAR} & \multicolumn{1}{c|}{BTC} & \multicolumn{1}{c|}{MONA} & \multicolumn{1}{c}{ETH} \\
        \hline
            & Minimum & 1.51 & 1.74 & 1.49 & 1.50 \\
            & Maximum & 4.22 & 10.99 & 7.69 & 3.42 \\
            \(\alpha\) & 1. Quartile & 1.94 & 2.42 & 2.46 & 1.68 \\
            & 3. Quartile & 2.32 & 2.93 & 3.21 & 1.81 \\
            & Median & 2.11 & 2.69 & 2.72 & 1.76 \\
    \end{tabular}
    \caption{Summary statistics on exponents of power-law fit of degree distribution on the weekly transaction networks.}
    \label{t:power-lawFit}
\end{table}

\begin{figure}[h]
    \centering
    \captionsetup{justification=centering}
    \includegraphics[scale=0.7]{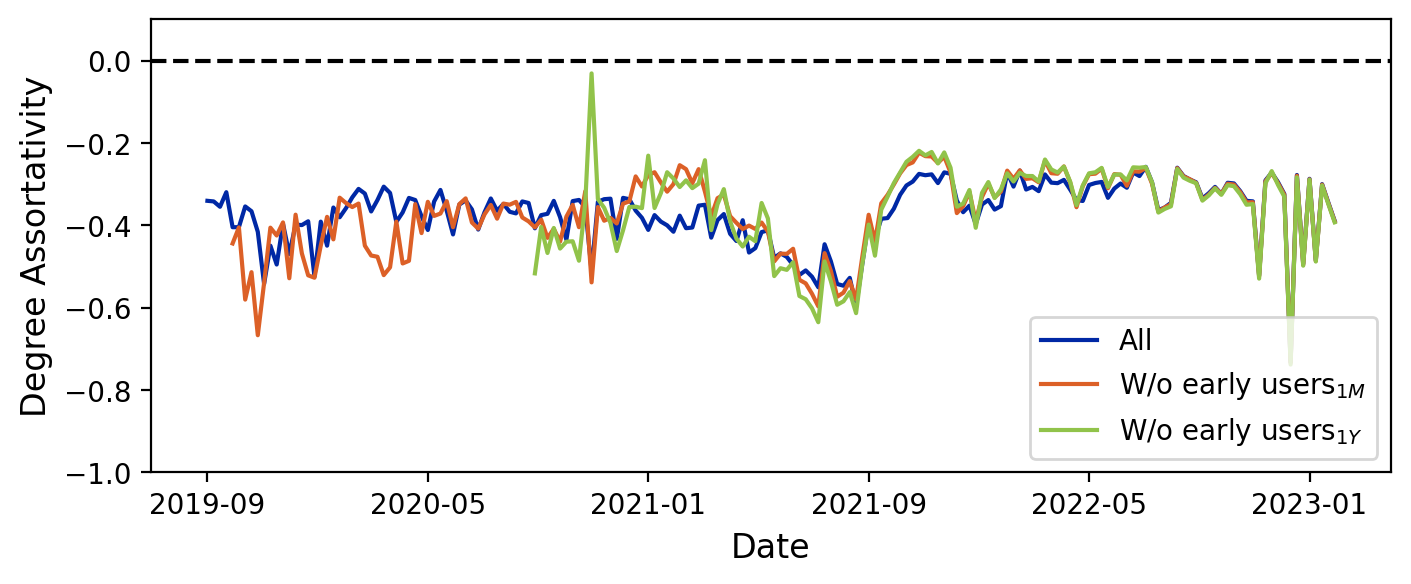}
    \caption{Assortativity coefficient for weekly transaction networks.}
    \label{fig:degree assortivity}
\end{figure}

A negative coefficient is found in figure \ref{fig:degree assortivity} for the entire period.
The disassortativity remains equal even if the accounts born in the first month/year are removed. Since the theoretical assortativity of an Erdős–Rényi (ER) random graph is zero, it can be stated that the weekly transaction networks we analyse do not fit the ER network model \cite{newman2002assortative}. Moreover, this result supports the view that the platform function similarly to centralised traditional financial markets, where many participants transact through a few dominant intermediaries. In these digital token economies, the intermediaries can take the form of exchanges, custodians or other service providers \cite{campajola2022evolution}.

\subsection{Core-Periphery Structure}\label{section:core periphery structure}

\begin{figure}[h]
    \centering
    \captionsetup{justification=centering}
    \includegraphics[scale=0.55]{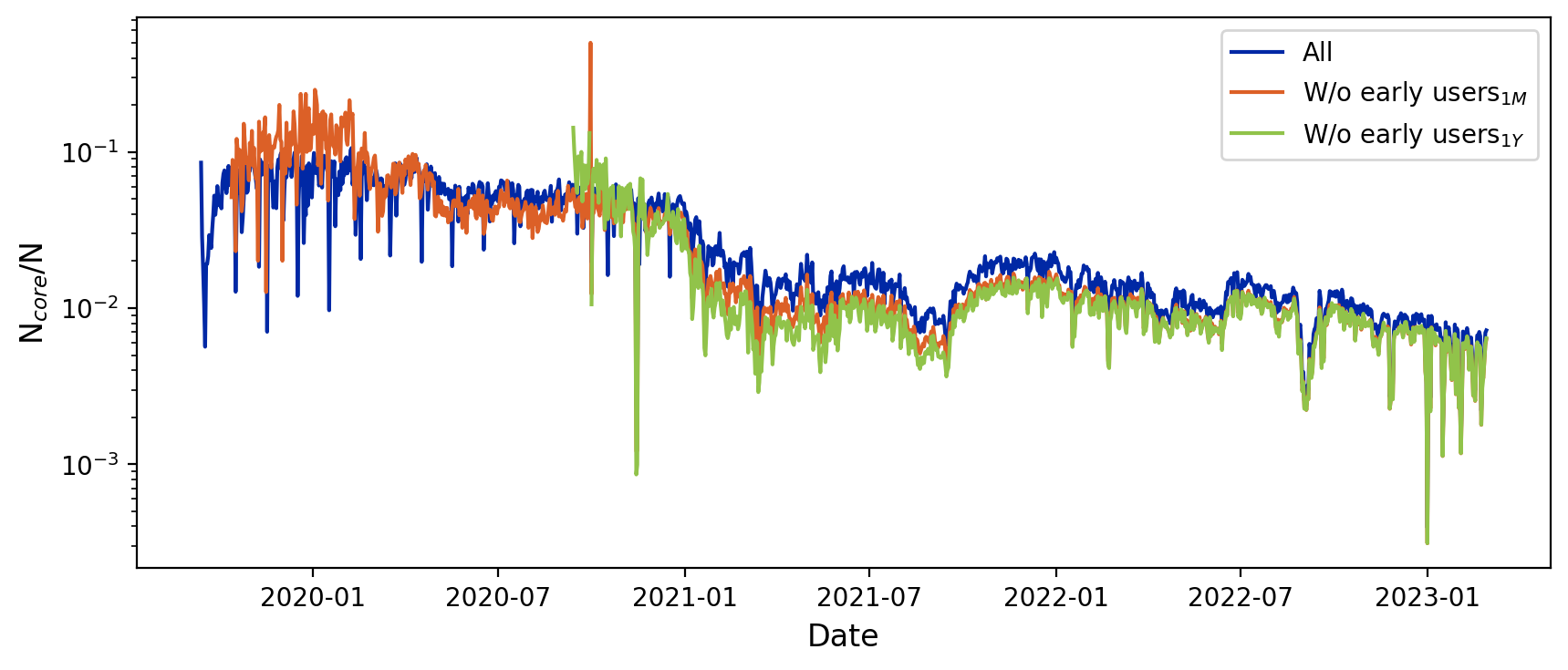}
    \caption{Number of core nodes as a fraction of all nodes.}
    \label{fig:core periphery ratio}
\end{figure} 

Here the algorithm suggested by Lip \cite{lip2011fast} is used for the bipartition of the actors and the relative size of the core is then shown as a fraction of all actors, in this case accounts, in figure \ref{fig:core periphery ratio}. In the early days of the networks, the core consists of a fraction of 5 - 10 \% of all actors and then starts to decline over time to a value lower than 1 \% at the end of Feb. 2023. If the early users (accounts whose birthday is in the first month or year of the analysed period) are removed, the core size does not change significantly compared to the core in the full network. Similar to what Campajola et al. \cite{campajola2022evolution} have found about other crypto economies, such as Bitcoin and Ethereum, this analysis finds a shrinking core over time, hinting at a more centralised transaction network over time and few actors that play the role of intermediaries, which makes this system not much different from the traditional financial system where the tokens are moved by a few big players that are sparsely connected to the majority of the network. \par

\subsection{Nakamoto Index}\label{section:mining centralisation}

\begin{figure}[h]
    \centering
    \captionsetup{justification=centering}
    \includegraphics[scale=0.7]{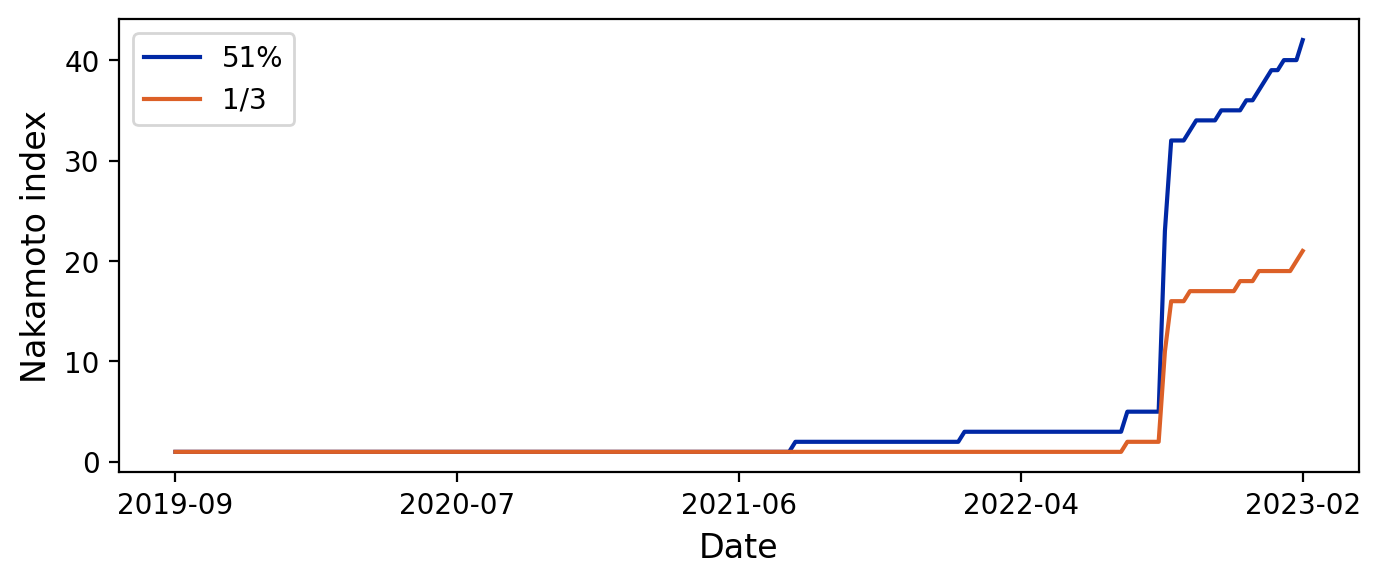}
    \caption{Hypothetical weekly Nakamoto index under the condition that the network was permissionless.}
    \label{fig:nakamoto index}
\end{figure}

Since the network is currently permissioned, i.e. only HGC members can run validator nodes, figure \ref{fig:nakamoto index} shows the hypothetical Nakamoto index for running a successful Sybil attack if the network was \emph{permissionless} and one stake in the balance represents one voting power. The 51\%-attack, a Sybil attack where 51\% of the voting power is required, is reported for comparability reasons since this is a commonly used threshold for other cryptocurrencies. The weekly Nakamoto indices are for almost three years since the launch of the network at 1, meaning one entity could have disrupted the network in its own interest, pointing to a very high centralisation even if the consensus is restricted to Council members. Only in September 2022 did it increase to more than 10 entities, starting an increasing trend. This compares to Nakamoto indices between 5 and 500 for Bitcoin, 1 and 20 for Monacoin and between 2 and 16 for Ethereum (all PoW-based) \cite{campajola2022evolution}. Overall, the results are striking suggesting that the PoS-based Hedera Hashgraph is even less decentralised by this gauge. Saad et al. \cite{saad2021pos} argue that this is no coincidence and is due to the consensus mechanism itself. \par 

Hence, it might prove valuable that Hedera takes this methodical approach to decentralisation. This permissioned approach is rationalised by Bakos, Halaburda, and Mueller-Bloch \cite{bakos2021permissioned} as they argue that it may not be intuitive at first but permissioned platforms are able to guarantee some level of decentralisation by binding many validators with off-blockchain contracts to participate in securing the system, which is not possible for permissionless systems where it is a result of free individual decisions. \newline


\subsection{Gini Coefficient}\label{section:gini coefficient}

\begin{figure}[h]
\centering
    \centering
        \captionsetup{justification=centering}
        \includegraphics[width=0.46\textwidth]{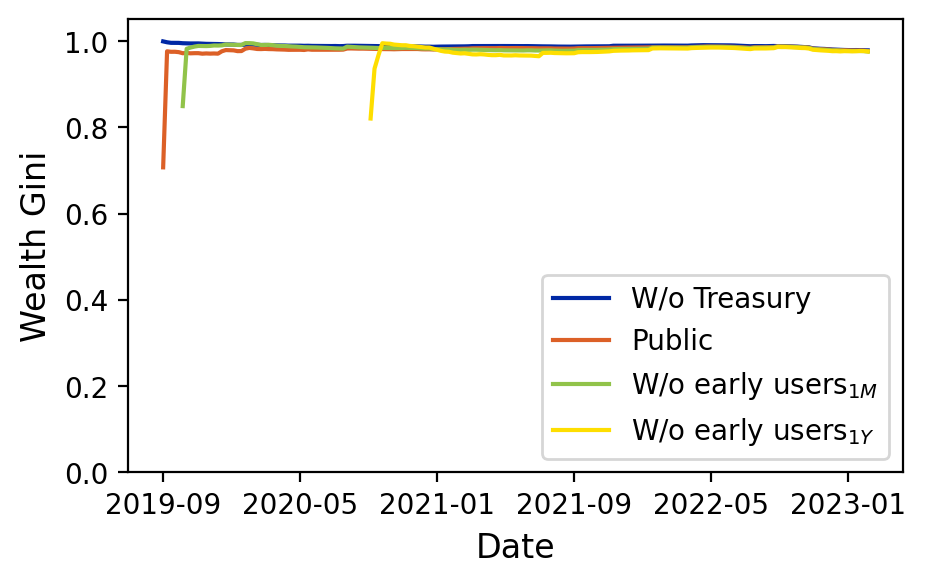}
        \includegraphics[width=0.46\textwidth]{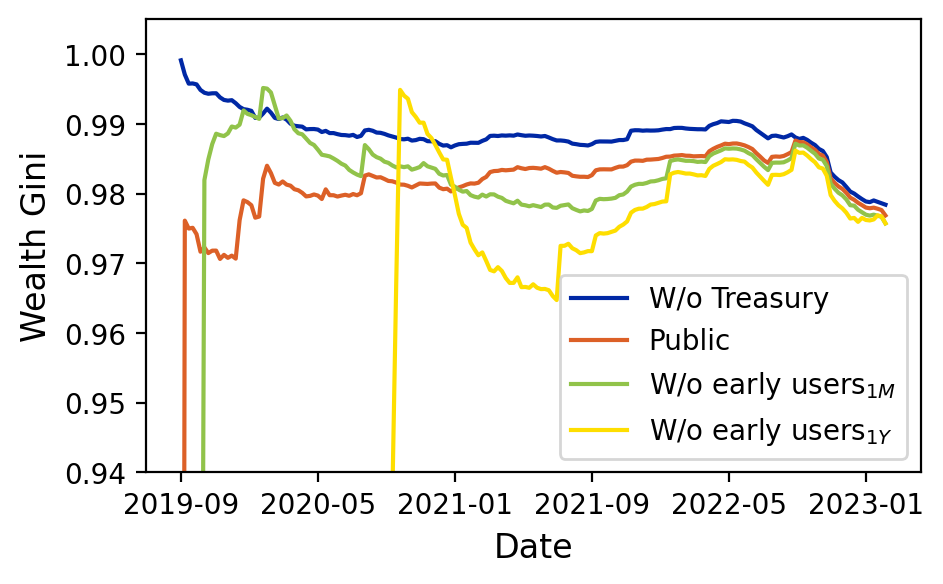}\label{fig:gini_zoom}
\caption{Gini coefficient of the balances - Accounts with a wealth of 10 HBAR or lower, the transaction fee account plus the staking reward account are excluded. In the right figure, the area of interest in y-axis is magnified.} \label{fig:gini_normal}
\end{figure}

The calculations in figure \ref{fig:gini_normal} show that the HBAR distribution on the network is almost perfectly unequal over the whole history. The label "Public" refers to all accounts but treasury and Hbar foundation accounts. Since all accounts with a wealth of 10 HBAR or less are excluded, the inequality is likely to stem from a few wealthy accounts. 
This result suggests a high wealth centralisation on Hedera Hashgraph and compares to very similar Gini coefficients found for Bitcoin, Ethereum, Dogecoin and others \cite{campajola2022evolution, gupta2018gini}. Another paper that merely considers the richest addresses in these blockchains finds lower values between 0.35 and 0.9 for Bitcoin, 0.45 and 0.65 for Ethereum and 0.55 and 0.85 for Dogecoin, though the small sample size of N=100 needs to be highlighted and the influence it can have on those low values \cite{kusmierz2022centralized}.

\subsection{Theil-T Index}\label{section:theil-t index}

\begin{figure}[h]
    \centering
    \captionsetup{justification=centering}
    \includegraphics[scale=0.5]{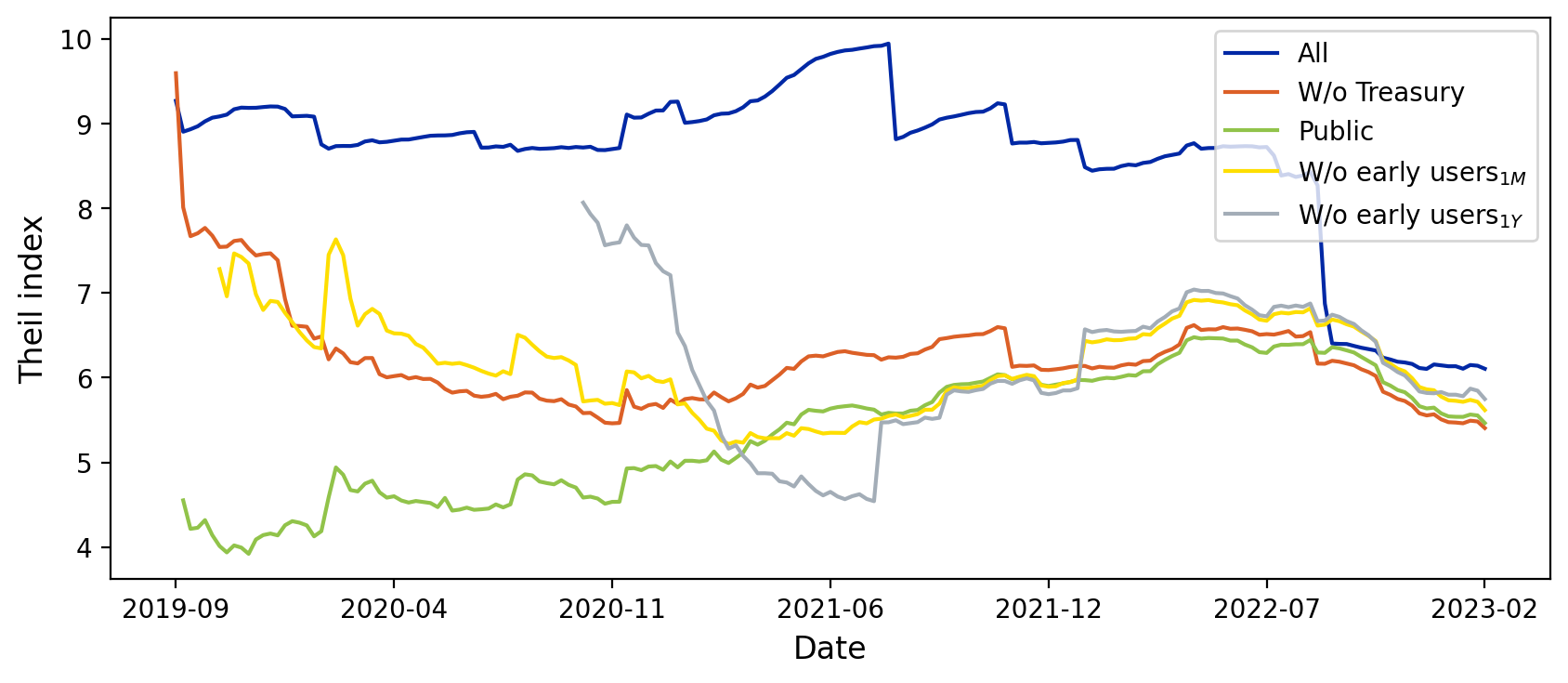}
    \caption{Theil-T index of the balances.}
    \label{fig:theil_bal}
\end{figure}

The weekly Theil-T index of HBAR holdings is shown in figure \ref{fig:theil_bal}. The results depict a weakly increasing inequality in the whole network until July 2021 and a clear trend to more equality of the distribution of wealth afterwards. For public accounts, there is a continuous growth in inequality until mid-2022. After that, the entire network trends to be more equal. Overall, one can find that there is a convergence of the Theil-T values of the different subgroups over time. 

\section{Discussion}\label{chapter:discussion}

This paper started by highlighting the importance of decentralisation in DLT systems, which can be seen as a reaction to the centralised state of the current financial system. However, in contrast to the common narrative, research literature finds that most of these DLT platforms exhibit a rather centralised state \cite{beikverdi2015trend,campajola2022evolution,gencer2018decentralization}. This paper adds to the literature by offering a first-time, data-driven analysis of the state of Hedera Hashgraph. This DLT system differentiates in several ways from other systems. Mainly, the data is not stored on the blockchain technology, but rather in a hashgraph. Additionally, the network is currently \emph{permissioned}, i.e. only known members of the Hedera Governance Council are allowed and required to run a validator node, with the objective of transitioning to a \emph{permissionless} system once it is enough decentralised. Due to these fundamental differences, the findings on the level of centralisation of Hedera Hashgraph provide a valuable extension for the existing literature. \par

In the first part of the analysis, the results of the growth of the network are discussed. On Hedera Hashgraph transactions employing the Hedera Consensus Service made up the overwhelming fraction of transactions from mid-2020, whereas transactions from the Hedera Token Service grew initially before plateauing in mid-2022. Looking at the correlation of network activity and HBAR returns, there is no correlation found, however, a strong positive correlation between the Bitcoin price and HBAR price is found. Moreover, the number of daily active accounts is growing faster over time, chiefly due to new active accounts not being born in the first year of the analysed period of the network, a sign of wider adoption of the network. Analysing the released supply, it is much higher than what is outlined by Hedera in their planning; for instance, 11.5 billion HBAR should have been released by the end of 2022 according to the documentation versus in reality 25 billion HBAR released \cite{hbareconomics}. The difference might come from selling pressure to finance the development of the platform and attract users, as it is unlikely that the supply was released due to faster progress on the path to decentralisation than scheduled.\par 

In the second part of the results, various gauges for the level of decentralisation are presented. We measure the Hedera decentralisation from perspectives of wealth distribution (using Gini and Nakamoto index) and token exchange (using degree assortativity and core-periphery structure). Striking is the Gini coefficient of the account balances, which remains close to 1 for the whole period, meaning that a very low number of accounts possess almost the entire wealth. 
In addition, a negative node degree assortativity and a shrinking core are found in the network, again, pointing to a centralised state of the transaction network, similar to a traditional financial market where many actors make transactions through only a few intermediaries such as exchanges that represent a kind of "interbank market". In a permissionless state of the network, Hedera Hashgraph has a Nakamoto index rapidly growing from 1 to over 10 at the end of 2022, starting an increasing trend towards more decentralisation. Similarly, the Theil-T index, a measure of economic inequality, displays a continuous increase in wealth equality in the entire network. \par
In conclusion, the analysis yields mixed signals on the level of decentralisation of Hedera Hashgraph. Gauges such as the Gini index, degree assortativity, and the core-periphery structure point towards a rather centralised state, but other metrics like the daily active accounts, the Nakamoto coefficient, and the Theil-T index yield progress towards decentralisation. Therefore, whether the permissioned approach to decentralisation on Hedera Hashgraph will lead to a better outcome in the future, remains to be seen.\par
Moreover, Hedera Hashgraph is an (almost) undiscovered land when it comes to research. Therefore, future research can develop in many different directions, for instance using more rigid analysis such as statistic tests, regressions, and machine learning. Building on this work, more granular network forensics can be done, especially further considering the known entities that were identified in this paper. Another extension towards a more comprehensive conclusion on the level of decentralisation of Hedera Hashgraph may analyse the staking mechanism and reward of the entities in the network. A third option to investigate the network in more detail is to evaluate the relationships between the accounts by creating an all-sided transaction network, also employing more methodologies from complex network science and knowledge about identifiable entities in order to extend the limited knowledge of real-world users of this study. \par

\ledgernotes









\bibliographystyle{ledgerbib}
\bibliography{tempbib}



\appendix
\setcounter{section}{0}



\section{Complementing figures}

\begin{figure}[h]
    \centering
    \captionsetup{justification=centering}
    \includegraphics[scale=0.45]{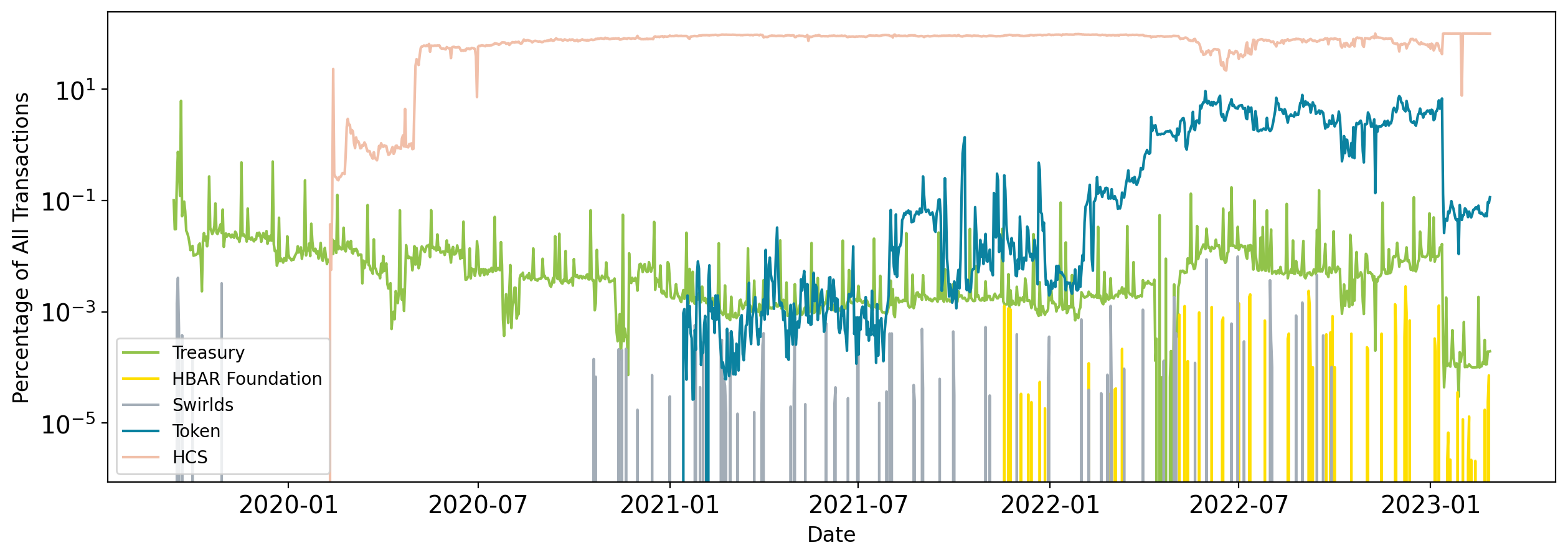}
    \caption{Number of transactions over time as a percentage of all successful transactions, categorized by type.}
    \label{fig:transaction_num_pct}
\end{figure}

\begin{figure}[h]
    \centering
    \captionsetup{justification=centering}
    \includegraphics[scale=0.45]{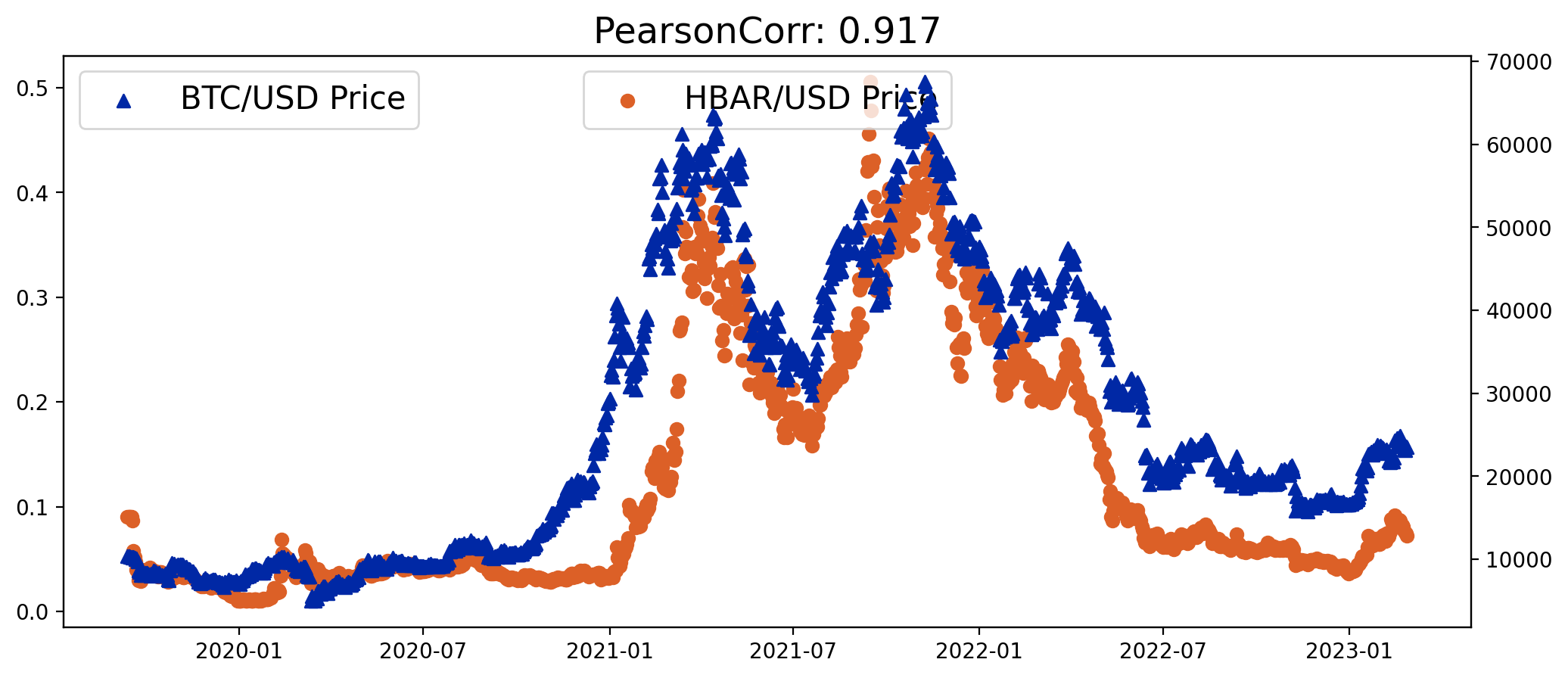} 
    \caption{Evolution of HBAR/USD price and BTC/HBAR price over time.}
    \label{fig:hbar-btc_corr}
\end{figure}

\thispagestyle{pagelast}


\end{document}